\documentclass[aps,twocolumn,prl,citeautoscript,showpacs,longbibliography]{revtex4}
\usepackage{graphicx}
\usepackage{color}
\usepackage{amsmath}
\usepackage{dcolumn}% Align table columns on decimal point
\usepackage{bm}% bold math

\begin{document}

\title{Spin-orbit  interactions in electronic structure quantum Monte Carlo}

\author{Cody A. Melton$^{1}$, Minyi Zhu$^{1}$, Shi Guo$^{1}$, Alberto Ambrosetti$^{2}$, Francesco Pederiva$^{3}$, Lubos Mitas$^{1}$}

\affiliation{
1) Department of Physics, North Carolina State University, Raleigh, North Carolina 27695-8202, USA\\
2) Dipartimento di Fisica, University of Padova, via Marzolo 8, I--35131, Padova, Italy\\
3) Dipartimento di Fisica e LISC, Universit\`a di Trento, Via Sommarive 14, I--38123, Povo, Trento,  and  Trento Institute for Fundamental Physics and Applications, Trento, Italy}

\date{\today}

\begin{abstract}
We develop generalization of the fixed-phase diffusion Monte Carlo
method for Hamiltonians which explicitly depend on particle spins such as for
 spin-orbit interactions. The method is formulated in zero variance manner and 
 is similar to treatment of nonlocal operators in commonly used static-spin calculations.
 Tests on atomic and molecular systems show that it is very accurate, on par with the fixed-node method. This opens electronic structure quantum Monte Carlo methods to a vast research area of 
quantum phenomena in which spin-related interactions play an important role.
\end{abstract}

\maketitle

%{\it Introduction.} 
 Quantum Monte Carlo (QMC) methods are making significant contributions
 to our understanding of many-body effects in quantum systems. 
  Although hampered by the infamous fermion sign problem,
  a number of approaches have been 
   explored for dealing with inefficiencies whenever sampled distributions possess varying
  signs or complex values. One of the commonly used
strategies is the fixed-node approximation that replaces 
the fermionic antisymmetry with boundaries given by trial wave 
function nodes. For broken time-reversal
 Hamiltonians or for twisted boundary conditions \cite{twisted} with inherently complex eigenstates,
 the fixed-node condition has been generalized to the fixed-phase approximation
\cite{ortiz}. 
  Benchmark quality results for both models and real materials have been 
  obtained in many settings
  such as molecules, solids, non-covalently bonded complexes, ultracold condensates and other
  systems \cite{qmcrev,qmcrpp}.
  
Electronic structure QMC calculations are usually done with particle spins being assigned fixed labels, up or down. 
Since spins commute with Hamiltonians without explicit spin terms, the problem simplifies to the spatial solution of the stationary Schr\"odinger equation.
Treating the spins as quantum variables for more complicated Hamiltonians was explored very early \cite{carlson} in variational Monte Carlo (VMC) of nuclear
systems. However, extending this  
to  projection methods such as the diffusion Monte Carlo (DMC) in position space \cite{qmcrev,qmcrpp} is much more involved.
 Building upon results for   
 nuclear systems  \cite{sarsa,pederiva, gandolfi}, a DMC method
 has been proposed and applied to
  a 2D electron gas with Rashba spin-orbit interaction  \cite{ambrosettidmc}.  
In this approach the spinors are stochastically updated by the action of the spin-orbit operator that is absorbed into the path sampling part of the propagator. It 
effectively samples the space of spinor states rather than (spin) coordinates and a similar VMC approach has been implemented for spin-orbit in atoms \cite{ambrosettivmc} very recently.

Here we propose a new development that is formulated as the DMC method in coordinate space with 
spinors in the trial state kept intact during the imaginary time evolution. This implies 
the zero variance property, ie, the bias in the  obtained energy is proportional to the square of the trial function error.

The method builds upon our previous work \cite{mitas} 
on nonlocal pseudopotentials (PP) since the spin-orbit operator 
is just another case
of inherent nonlocality. 
It is also well suited for calculations of real systems with
heavy atoms since both scalar relativistic and spin-orbit effects 
can be accurately represented by pseudopotentials
 as it is routinely done in quantum chemical calculations \cite{DolgChemRev}.
In particular, commonly used semilocal one-particle PP operator
$W=\sum_{l,m}v_l(r)| lm\rangle \langle  lm|$ is generalized to
\begin{equation}
W= \sum_{l,j} v_{lj}(r) \sum_{m_j}| ljm_j\rangle \langle  ljm_j|
\end{equation}
where $| ljm_j\rangle$ are atomic one-particle spinors, $v_{lj}$ 
are potential functions while $r$ is the distance from the ion. The method employs the fixed-phase approximation and 
 therefore depends on the accuracy of the trial function similarly to the fixed-node DMC with static spins. 

{\it Phase and absolute value.} We assume a Hamiltonian $H=T+V+W$, 
where $V$ denotes electronic and ionic local interactions while
 $W$ represents nonlocal and spin-orbit PP terms. 
 Substituting $\Psi=\rho\exp(i\Phi)$ into the imaginary-time 
 Schr\"odinger equation gives for the real component
\begin{equation}
-\partial_t \rho = [T+V+W^{Re}+(\nabla\Phi)^2/2]\rho
\end{equation}
where $W^{Re}={\rm Re}[\rho^{-1}\exp(-i\Phi)W\rho\exp(i\Phi)]$.
The imaginary component equation describes the phase $\Phi$ flow conservation
between sources and sinks represented 
by the imaginary part of $W$; however, it has no contribution 
to the total energy since $W$ is Hermitian.
In the limit $t\to \infty$, the stationary solution of the real part provides
the desired ground state energy and corresponding $\rho$. 

{\it Approximations.} In general, neither the exact
phase $\Phi$ nor $W^{Re}$ are known
and we have to introduce approximations. 
First, we impose the fixed-phase approximation in which 
$\Phi$ is replaced by the trial wave function phase.
The corresponding potential is given as 
\begin{equation}
(\nabla\Phi)^2/2 \approx  (\nabla\Phi_T)^2/2
\end{equation}
For a given $V+W^{Re}$ in an ordinary representation it follows that
this approximation is variational \cite{ortiz},
ie, the energy expectation with $\rho\exp(i\Phi_T)$
is an upper bound to the exact value for
an arbitrary symmetric $\rho\geq 0$. 
Since the fixed-phase solution for $\rho$  
is nonnegative everywhere by construction, the fermion sign problem is eliminated variationally; this implies the need for accurate approximations to the many-body phase. For the sake of completeness we also note that the fixed-node approximation commonly used in QMC with real wave functions is a special case of the fixed-phase approximation, as had been pointed out by Ortiz {\it et al} \cite{ortiz}. In addition, we note that in
twist-averaging (Brillouin zone sampling) calculations of periodic systems one employs both
fixed-node and fixed-phase approximations on equal footing
 since they typically exhibit comparable systematic and statistical errors \cite{twisted}.
 
The second
approximation involves the projection of $W$ onto the trial 
function, similarly to the localization approximation %approximation
for nonlocal pseudopotentials \cite{mitas}   
in spin-free Hamiltonians. $W^{Re}$ is approximated as 
\begin{equation}
W^{Re}  \approx  W^{Re}_T={\rm Re} [\Psi^{-1}_TW\Psi_T].
\end{equation}
This results in a multiplicative many-body operator and
one can show that the bias in energy obtained with 
$W^{Re}_T$ vanishes quadratically
in the trial function error \cite{mitas}. The resulting energy 
is not necessarily variational, however, the variational property
can be recovered with an appropriate modification \cite{melton2}
of the T-moves algorithm \cite{casula}. 

{\it Continuous spin and its updates.}
 In its usual representation, the spin configurations have
discrete values  $\pm 1/2$ so that for $S_z$ eigenstates 
 $\chi^{\uparrow}(1/2)=\chi^{\downarrow}(-1/2)=1$,
$\chi^{\downarrow}(1/2)=\chi^{\uparrow}(-1/2)=0$. The corresponding 
 configuration space is non-compact and therefore
would lead to ``jumps" in the evolving stochastic paths. 
Such jumps could easily make the sampling process rather inefficient since the corresponding local energy fluctuations could go up substantially. Large fluctuations would complicate both the importance sampling and make calculations of larger systems intractable and, eventually, unreachable. Another straightforward option would be to integrate over all spins for every spatial step, however, this would scale exponentially due to the  $2^N$ configurations for $N$ electrons.
One way to overcome this difficulty is to
introduce a continuous, {\em overcomplete}
and compact representation \cite{kevin}, that 
enables to make the paths smooth.
Obviously, we also wish
that the spin coordinate space is small so that
the sampling can be fast. 
This points towards the pair of orthogonal states
for a 1D ring as one possible option 
\begin{equation}
\langle s_i|\chi^{\uparrow}\rangle=e^{is_i};\;
\langle s_i|\chi^{\downarrow}\rangle=e^{-is_i};\; 
\langle \chi^{\alpha}|\chi^{\beta}\rangle=2\pi\delta_{\alpha\beta}.
\end{equation}
We note that overcompleteness can also compromise the variational property although we estimate that the dominant source of such possible bias would be the localization approximation.

The sampling of spins is done in a manner similar to the spatial degrees 
of freedom. For this purpose we add spin
``kinetic" energies into the Hamiltonian $H$ for all $s_i, i=1,...,N$.
It includes an effective mass $\mu_s$  and an energy offset
\begin{equation}
T_{s_i}= - {1\over 2\mu_s}\left[{\partial^2 \over \partial s_i^2}+1\right]
\end{equation}
so that it annihilates an
 arbitrary spinor  $\psi$
 \begin{equation}
 T_{s_i}\psi=T_{s_i}[\alpha\varphi^{\uparrow}({\bf r}_i)\chi^{\uparrow}(s_i)
 +\beta\varphi^{\downarrow}({\bf r}_i)\chi^{\downarrow}(s_i)]=0.
\end{equation}
The offset cancels out the bare spin kinetic contribution; 
however,
$T^s$ does not commute with $H$ so there is some additional contribution to the energy.
For the considered strengths of spin-orbit,
this contribution appears to be small and can be fully eliminated by running the effective spin mass $\mu_s$ to zero; this effectively increases the corresponding 
diffusion constant and in turn speeds up the spin sampling. Another significant effect of such faster spin evolution is that a subset of possible spin configurations gets sampled per single spatial step. This partial averaging statistically approximates the full average over the $2^N$ space. Since one can adjust the spatial and spin time steps independently, it is possible to carry out extrapolations to find the unbiased values. 
Due to the fact that the spin functions are very smooth, in the tested cases
we found that the spin time step that is 5-10 times larger than the spatial one was sufficient such that the resulting energies were not affected.
 
{\em Importance sampling and trial function.} 
The final step is the importance sampling that is accomplished by multiplying the real part of the Schrodinger equation with the trial function $\rho_T$. The trial function is
a product of the Jastrow factor and Slater determinant(s) of spinors 
\begin{equation}
\Psi_T({\bf R,S}) = \exp[U({\bf R})] \sum_k c_k {\rm det}_k [\{\psi_n ({\bf r}_i,s_i) \}]
\end{equation}
where ${\bf R}=({\bf r}_1,...,{\bf r}_N)$ and ${\bf S}=(s_1, ...,s_N)$.
The Jastrow factor includes electron-ion, electron-electron and, possibly, higher order terms.  
 Since $U({\bf R})$ depends only on the spatial 
 coordinates, the spin integrations
in the nonlocal operator can be done explicitly and the rest is similar
to the treatment of nonlocality
in static spin calculations \cite{mitas,melton2}. 
The short-time approximation for the importance sampled propagator \cite{qmcrev,qmcrpp} is a product of the dynamical and reweighting factors $G({\bf R,S; R',S'})$ $=$ $G_{dyn}e^{-\Delta t(E_{loc}+E'_{loc}-2E_T)/2}$, where $E_{loc}({\bf R,S})=[H\Psi_T]/\Psi_T$. 
Note that as $\Psi_T$ converges to the exact eigenstate the local energy approaches the exact eigenvalue
with vanishing variance pointwise, regardless of the representation or the propagator accuracy. Note that the success of the method depends on the local energies to be mildly varying since large fluctuations could cause very large variance of the exponentials and make the sampling very inefficient. More details about this fixed-phase spin-orbit DMC (FPSODMC) method are further elaborated elsewhere \cite{melton2}.

 \begin{figure}[!b]
\centering
\caption{Total energies of the lowest states of Pb atom from the FCI  method (circles)
with cc-pV$n$Z basis sets compared to FPSODMC with a COSCI trial wave function (dashed lines).  
The valence space includes only $6s$ and $6p$ electrons.
\\}
\includegraphics[width=0.485\textwidth] {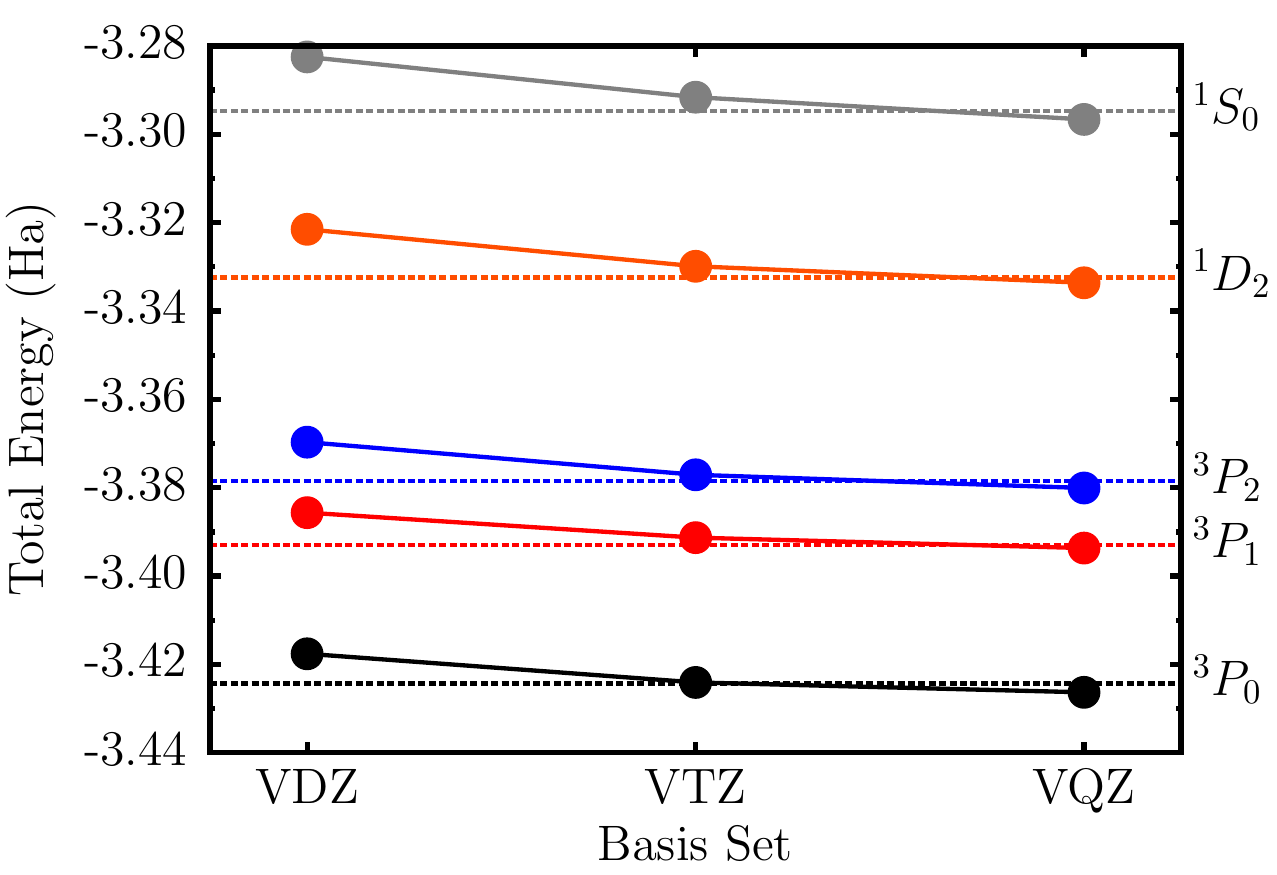}
\label{pbtotal}
\end{figure}

\begin{table}[!t]
\centering                                      % used for centering table
\caption{FPSODMC (DMC for short) excitation energies [eV] of the Pb atom 
using LC and 
SC relativistic PPs \cite{STU2000} with COSCI trial function compared with experiment (Exp). For completeness we include
also Dirac-Fock COSCI results. The multi-reference CI (MRCI) and MRCI + core polarization corrections (CPP)
calculations \cite{STU2000} are done with the LC PP.} 
% title of Table
\renewcommand{\arraystretch}{1.2} 
\begin{tabular}{c c  c c c c c c r}
\hline\hline
           &$\;\;$COSCI$\;\;$  & $\;\;$ DMC $\;\;$ &$\;\;$DMC$\;\;$ &MRCI &MRCI\quad &Exp&\\[0.5ex]
State   &LC & LC&SC &  &   +CPP &\cite{NIST} \\
\hline % inserts single horizontal line
$^3P_1$ &0.83  &0.851(1) & 0.90(1) &0.90 &0.94& 0.97 \\
$^3P_2$  &1.30  &1.245(4) & 1.10(1) &1.27 &1.32& 1.32 \\
$^1D_2$  &2.69  &2.500(4) & 2.42(1) &2.55 &2.66& 2.66 \\
$^1S_0$  &4.06  &3.527(5) & 3.42(1) &3.54 &3.68 & 3.65 \\[0.5ex]\hline
\end{tabular}
\label{table:Pb_lcsc}
\end{table}

{\em Atomic calculations: excitations in Pb, Bi, and W atoms.} 
We present results for the lowest excitations of Pb, Bi and W atoms as a testing ground for atomic calculations  
  with spin-orbit effects.  
  This choice is motivated by several considerations. Clearly, the
  spin-orbit interaction is appreciably large in Pb and Bi. These atoms are often used as the simplest illustrations of the 
  spin-orbit splittings since they exhibit an open shell with only two  and three $p$ states, respectively.  That makes the splittings at the linear
  combination of orbitals level analytically transparent and is often used in textbooks \cite{gottfried}. At the same time, the spin-orbit in these atoms have impact beyond just finding the corresponding multiplet energies 
  for the well-understood cases. In particular, spin-orbit induced shift in the ground state energy changes the key quantities such as energies in chemical bonds by very large amounts as we illustrate later.
  
  The W atom calculations illustrate another important point.
  As demonstrated below, the interplay of spin-orbit and correlation is needed to find basic properties such as the symmetry and occupation in the ground state of this atom. 
  Note that the effect is significant since these properties 
  are different from its isovalent elements in the same column
  of the periodic table such as Cr and Mo, despite the fact that Mo exhibits a sizeable spin-orbit interaction as well. Therefore the example of W atom is quite revealing as a
  demonstration of these combination of effects.
  
  The chosen examples show the introduced method to be on par with essentially the only total energy, wave function-based alternative, namely, the  expansions in Slater determinants with large basis sets. 
  In some cases, the accuracy and agreement between the two approaches enabled us to reveal the accuracy limits of the existing PPs and to point out that a new generation of 
  PPs will be needed in order 
  to harness the full potential of such accurate QMC calculations. Note that the used PPs were constructed in a Dirac-Hartree-Fock formulation, therefore one does not 
expect their accuracy to be systematically better than 0.1-0.2 eV
  for energy differences.
  Additionally, QMC is scalable to much larger systems. Clearly this positions our method for promising prospects for high-accuracy
  correlated calculations of larger molecular and solid systems with spin-dependent 
  Hamiltonians.
  
  In the valence-only framework 
we have tested two types of relativistic PPs for Pb and Bi with spin-orbit terms: large-core (LC) with 4 and 5 valence 
electrons and small-core (SC) with 22 and 23 valence electrons respectively \cite{STU2000,STU2002}. 
 The spinors for the Slater determinants are obtained from the 2-component Dirac-Fock with complete open-shell configuration interaction (COSCI) calculations using an extensive basis set and the 
 Dirac13~\cite{Dirac13} code. 
 For the COSCI trial wave functions, we include only the local, $s$, $p_{3/2}$, and $p_{1/2}$ channels in the LC PP, for the SC we add  $d_{5/2}$ and $d_{3/2}$.  
 
 \begin{figure}[!b]
\centering
\caption{Total energies of the lowest states of Bi atom from CISDT (circles) with cc-pV$n$Z basis sets
compared to FPSODMC with COSCI (long-dashed lines) and 
CISDT (short-dashed lines) trial wave functions.  The valence space 
includes only $6s,6p$ states. 
}
\includegraphics[width=0.485\textwidth] {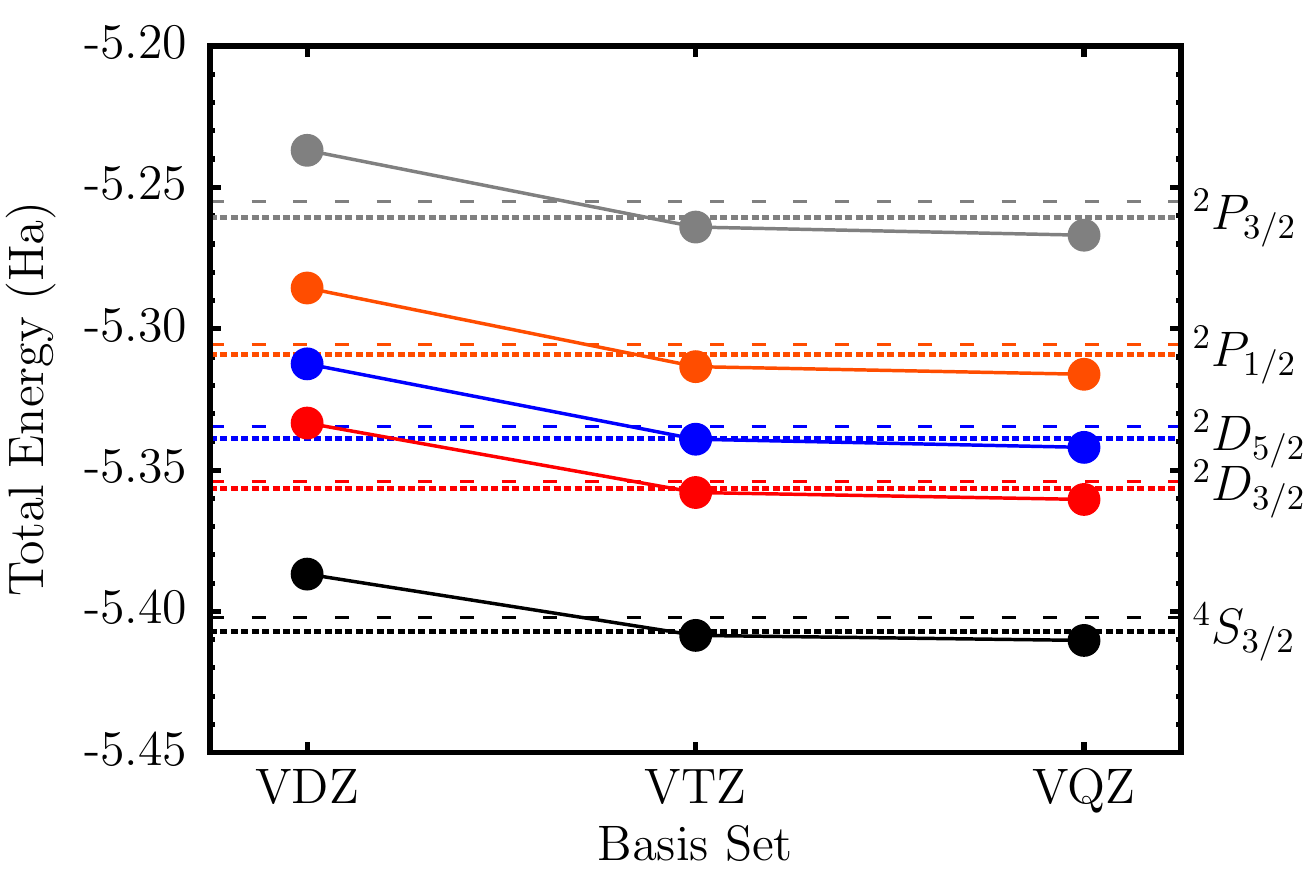}
\label{bitotal}
\end{figure}

 We note that unlike in 
 $ls$-coupling
 the only good quantum number is the total angular momentum $J$ and 
 the lower symmetry is indeed manifested 
 in significant mixing of the states. For example, the ground state $^3P_0(6s^26p^2)$ mixes very strongly with the highest state $^1S_0(6s^26p^2)$ with an amplitude of $\approx 0.2$.
Using the same Hamiltonian, we perform a full configuration interaction (FCI) in the two-component formalism to compare with our fixed-phase spin-orbit DMC (FPSODMC) calculations. 
The total energies from FPSODMC and FCI are shown in Fig.~\ref{pbtotal}. The agreement between the total energies is excellent showing that as soon as the most 
significant configurations are included, the fixed-phase approximation shows accuracy 
that is similar to the conventional static-spin calculations in the fixed-node approximation.
 The energy differences are listed in Tab. \ref{table:Pb_lcsc} and agreement with the experiment
 is very good although not perfect, due to small biases in the 
 PP that are visible from essentially perfect agreement between DMC and FCI methods. In addition, we calculate the electron affinity (EA) for the Pb atom that is significantly lower than for other group IVB elements due to the spin-orbit interaction.  
 We find the EA to be 0.417(7)~eV that, within the accuracy of the PPs, compares very favorably to the experimental value 0.365(8)~eV \cite{eaPbExpt} and also to other theoretical values \cite{eaPb}.  
In SC calculations the first excited state is closer to experiment than the LC, however the higher excitations are off by $\approx$~0.2~eV that we assign mostly to the used PP imperfections. 

 For Bi we calculate the first four excitations of the $p^3$ electronic configuration and for an independent validation
we perform CI with single, double, and triple (CISDT) excitations.
Results are shown in Fig. \ref{bitotal}. The FPSODMC using a COSCI wave function agrees with the total energies for the CISDT to only $\approx$~.007~Ha. 
We also test the improvement of the fixed-phase error by including higher excitations into virtual spinors with resulting closer agreement with the CISDT energies. 
Although the COSCI is missing some correlation energy (approx. 5-9\%) compared to the more extensive FPSODMC/CISDT and CISDT methods, it only adds a constant shift to the spectra since the excitation energies are comparable between all the correlated methods in Tab. \ref{table:Bi_lcsc}. 
We see good agreement for both the LC and SC calculations with the COSCI trial wave function. The FPSODMC with the CISDT trial wave function most accurately reproduces the experimental values.

\begin{table}[!t]
\centering                                      % used for centering table
\caption{FPSODMC excitation energies [eV] of the Bi atom 
using LC and 
    SC relativistic PPs \cite{STU2000,STU2002} compared with experiment (Exp). For completeness we 
    include Dirac-Fock complete open-shell CI (COSCI), CI (SDT) results. The second row indicates FPSODMC trial wave functions. 
} 
% title of Table
\renewcommand{\arraystretch}{1.2} 
\begin{tabular}{c | c c c |c c | c }
\hline\hline
           &$\;\;$COSCI$\;\;$  & $\;\;$ DMC/ $\;\;$ &$\;\;$DMC/$\;\;$ & CI&  DMC/ &Exp\\[0.05ex]
  &  & COSCI &COSCI&  & CI&  \\[0.05ex]
State			      & LC & LC&SC & LC & LC & \cite{NIST} \\
\hline  
$^2D_{3/2}$ & 1.542 & 1.311(4) & 1.38(1) & 1.356 & 1.37(2) & 1.415 \\
$^2D_{5/2}$ & 2.129 & 1.834(6) & 1.74(1) & 1.858 & 1.85(2) & 1.914 \\
$^2P_{1/2}$ & 3.108 & 2.628(6) & 2.53(1) & 2.562 & 2.66(2) & 2.685 \\
$^2P_{3/2}$ & 4.428 & 4.005(6) & 3.95(1) & 3.900 & 3.98(2) & 4.040 \\[0.25ex]\hline
\end{tabular}
\label{table:Bi_lcsc}
\end{table}

\begin{table}[!t]
\centering                                      % used for centering table
\caption{DMC excitation energies [eV] of the W atom with a 
    relativistic PP \cite{STU2009} compared with CISD and experiment (Exp). 
 CISD is extrapolated to a complete basis set limit.
} 
% title of Table
\renewcommand{\arraystretch}{1.2} 
\begin{tabular}{c c c c c c c c}
\hline\hline
Config. & State & COSCI  &  DMC/   &  CISD  &    DMC/   & Exp\\
              &       &     &    COSCI  &          &    CISD & \cite{NIST} \\
\hline % inserts single horizontal line
$5d^46s^2$ & $^5D_1$ &  0.098 &  0.130(9) & 0.104 & 0.15(1) & 0.207\\
$5d^56s^1$ & $^7S_3$ & -0.845 & -0.194(9) & 0.115 & 0.19(1) & 0.365\\
$5d^46s^2$ & $^5D_2$ &  0.244 &  0.30(1)  & 0.132 & 0.30(1) & 0.412\\
$5d^46s^2$ & $^5D_3$ &  0.415 &  0.49(1)  & 0.289 & 0.51(1) & 0.598\\
$5d^46s^2$ & $^5D_4$ &  0.599 &  0.686(9) & 0.452 & 0.69(1)& 0.771\\\hline
\end{tabular}
\label{table:W}
\end{table}

The next system we calculate is W atom that shows the importance 
of both spin-orbit as well as the electron correlation.
It is an interesting case since the isovalent Cr and Mo atoms
have the ground state occupations $d^5s^1$ whereas the ground state of W is $d^4s^2$. Qualitatively, the $d^4$ occupation is favored due to the lower energy 
in the $j=3/2$ channel, however, it turns out that correlations have to be captured accurately as well. 
We used a relativistic PP with 14 electrons \cite{STU2009} with two different trial wave functions, COSCI and CISD.
The results are listed in Tab. \ref{table:W}. 
Clearly, the ground state of the Dirac-Fock COSCI method is $5d^56s^1$, indicating that correlation is crucial for correctly calculating the spectrum. 
Using the CISD as a trial wave function in FPSODMC, we not only see the states are correctly ordered, but the excitation energies are accurate to within $\approx$~0.1~eV.  We note that the
FPSODMC/CISD energies are significantly lower than the ones from the basis set extrapolated CISD. 

\begin{table}[!b]
        \centering
        \normalsize
        \setlength{\tabcolsep}{2pt}
        \caption{PbO bond length ($r_e$) and dissociation energy ($D_e$). The DMC
        calculations are done with the small-core PP and averaged SO represents the fixed-node DMC.}
  \renewcommand{\arraystretch}{1.1}
\begin{tabular}{ l  c  c }
\hline\hline
\qquad  Method     &\qquad $r_e$ (\AA)      &$D_e$ (eV)   \\[0.5ex] % inserts table heading
\hline % inserts single horizontal line
~spin-free PP-CCSD(T)$ ^a$ &\qquad 1.886 &5.14\\
~MRCIS-spss/CCSD(T)$ ^b$  &\qquad1.923 & 3.87\\
~spin-free AE-CCSD(T)$ ^c$ &\qquad1.937 &4.85\\
~SO AE-CCSD(T)$ ^d$     &\qquad1.934 &3.63\\[0.1ex]
 ~$ \text{DMC/one-comp. averaged SO}$  &\qquad$\;\;\;$1.88(1) &$\;\;\;\,$ 4.76(3)\\
 ~$ \text{FPSODMC/2-comp.}$ &\qquad$\;\;\;$1.92(1) & $\;\;\;\,$ 3.83(3)\\
 ~Exp.$^e$ & \qquad 1.920  &  3.87 \\
\hline
 \multicolumn{3}{l}{ \small{~$^a$  spin-free CCSD(T) one-comp., averaged SO PP \cite{STU2000} }}\\[-1mm]
\multicolumn{3}{l}{ \small{~$^b$ 2-comp.  MRCIS with spin-free-state shift evaluated  }}\\[-1mm]
\multicolumn{3}{l}{ \small{~~~~~~~~~~~~~~ with one-component CCSD(T)\cite{STU2000}}}\\[-1mm]
 \multicolumn{3}{l}{ \small{~$^c$ all-electron spin-free CCSD(T) \cite{AEPBO}  }}\\[-1mm]
 \multicolumn{3}{l}{ \small{~$^d$ all-electron  SO CCSD(T) \cite{AEPBO}   }}\\[-1mm]
 \multicolumn{3}{l}{ \small{~$^e$ Experimental data \cite{exp_PbH/PbO}  }}
        \end{tabular}
 \label{table:PbO}
\end{table}

{\em Molecular calculations.} For the PbO molecule
we use the SC PP as recommended \cite{STU2000} so as to avoid 
overlaps between PPs from the two atoms. 
  The theoretical results of  bond length and dissociation energy of PbO molecule together with  experimental
   values are given in Table \ref{table:PbO}.
 We also report 1-component CCSD(T) combined with 2-component MRCI studies 
 \cite{STU2000, AEPBO} both in PP and all-electron, frozen-core
 settings.  
The  bond length $r_e$  with the SO interaction shows an excellent agreement with experiment value, compared to an underestimation by $\tiny \sim$0.04 \AA ~ in static-spin PP calculations.  We also note that the averaged SO treatment overestimates the dissociation energy by $\approx$~0.9~eV whereas we see excellent agreement by explicit treatment of the SO effects.

In conclusion, we have proposed a new projector QMC method for treating the spins as quantum variables in electronic
structure calculations. 
The method establishes continuous spin coordinate sampling with resulting zero variance algorithm within the fixed-phase approximation and projections of the nonlocal pseudopotentials. The tests on atomic and molecular systems for both total energies and differences show excellent agreement with independent correlated quantum chemical calculations in two-component formalism. The accuracy is very similar to the fixed-node DMC that is widely used for static spins calculations. The method opens QMC to 
variety of systems across the periodic table such as materials with non-collinear states, spin waves and other electronic phases for which particle spins are of the key importance.

{\em Acknowledgments.}
Discussions with 
J. Koloren\v c  and R. Derian are gratefully acknowledged. 
Major part of this research was supported by the U.S. Department of Energy (DOE), Office of Science, Basic Energy Sciences (BES) under Award de-sc0012314. 
This research used resources of the National Energy Research
Scientific Computing Center, a DOE Office of Science User Facility 
supported by the Office of Science of the U.S. Department of Energy 
under Contract No. DE-AC02-05CH11231.

\end{document}